\documentclass[smallextended]{svjour3}     
\smartqed  
\usepackage{graphicx}
\usepackage{natbib}
\usepackage[innercaption]{sidecap}
\usepackage{mathptmx}      
\newcommand{\aj}{AJ, }
\newcommand{\araa}{ARA\&A, }
\newcommand{\apj}{ApJ, }%
\newcommand{\apjl}{ApJ, }
\newcommand{\ao}{Appl.~Opt, }%
\newcommand{\aap}{A\&A, }
\newcommand{\aapr}{A\&A~Rev., }%
\newcommand{\aaps}{A\&AS, }%
\newcommand{\mnras}{MNRAS, }%
\begin{document}

\title{A new concept for the combination of optical interferometers and high-resolution spectrographs
}

\titlerunning{Combination of optical interferometers and high-resolution spectrographs}        

\author{S. Albrecht \and A. Quirrenbach \and R. N. Tubbs \and R. Vink }

\authorrunning{Albrecht et al.} 

\institute{S. Albrecht \at
  Leiden Observatory, Leiden University,  P.O.\ Box 9513, NL-2300 RA  Leiden, The Netherlands \emph{Present address: MIT, Kavli Institute for Astrophysics and Space Research, Cambridge, Massachusetts 02139, USA}
  \email{albrecht@space.mit.edu}
  \and A. Quirrenbach
  \at  Landessternwarte, K\"{o}nigstuhl 12, D-69117 Heidelberg, Germany
  \and R. N. Tubbs
  \at Max-Planck-Institut f\"{u}r Astronomie, K\"{o}nigstuhl 17, D-69117 Heidelberg, Germany \emph{Present address: Met Office, PO Box 243, READING RG6 6BB, United Kingdom}
  \and  R. Vink
  \at European Space Agency, ESTEC, Keplerlaan 1, NL-2200 AG Noordwijk, The Netherlands
}

\date{Received: date / Accepted: date}

\maketitle

\begin{abstract}

The combination of high spatial and spectral resolution in optical
astronomy enables new observational approaches to many open problems
in stellar and circumstellar astrophysics. However, constructing a
high-resolution spectrograph for an interferometer is a costly and
time-intensive undertaking.  Our aim is to show that, by coupling
existing high-resolution spectrographs to existing interferometers,
one could observe in the domain of high spectral and spatial
resolution, and avoid the construction of a new complex and expensive
instrument. We investigate in this article the different challenges
which arise from combining an interferometer with a high-resolution
spectrograph. The requirements for the different sub-systems are
determined, with special attention given to the problems of fringe
tracking and dispersion. A concept study for the combination of the
VLTI (Very Large Telescope Interferometer) with UVES (UV-Visual
Echelle Spectrograph) is carried out, and several other specific
instrument pairings are discussed.  We show that the proposed
combination of an interferometer with a high-resolution spectrograph
is indeed feasible with current technology, for a fraction of the cost
of building a whole new spectrograph. The impact on the existing
instruments and their ongoing programs would be minimal.

\end{abstract}

\section{Introduction}

In recent years optical interferometers have proven that they can
produce excellent science in the field of stellar and circumstellar
astrophysics. Over the same period high-resolution spectrographs have
enabled the discovery of the first extra-solar planets, and
contributed substantially to great progress in the field of
asteroseismology.

A number of current interferometric instruments have some
spectroscopic capabilities. For example the mid- and near-infrared
instruments MIDI (The Mid-Infrared instrument, at the VLTI) and AMBER
(Astronomical Multiple BEam Recombiner, at the VLTI) provide spectral
resolutions of up to $R\sim250$ and $R\sim12 000$, over
bandpasses of $\sim 5$~$\mu$m and $\sim 50$~nm, respectively.
At the CHARA array, the Vega (Visible spEctroGraph and polArimeter)
project is under construction with a spectral resolution of
$R\sim30 000$ and a bandpass of $\sim 50$~nm.  For science
results obtained with spectrally resolved interferometry see for
example \cite{vakili1998} and \cite{weigelt2007}. Unfortunately the
combination of very high spectral resolution over a bandpass greater
than a few tens of nanometer, to enable a real analog to classical
Echelle spectroscopy with interferometric spatial resolution is not
yet available.

In building a dedicated high-resolution Echelle spectrograph for an
existing interferometer such as the VLTI, one would face several
challenges. First of all, building a high-resolution spectrograph is a
very costly and time-intensive undertaking. In addition it would be
hard to justify building such an instrument only for use with an
interferometer, as current optical interferometers are still
restricted to very bright objects in comparison with single
telescopes. Furthermore, high-resolution spectrographs are usually
large instruments, while the space available in the beam combining
laboratories of interferometers is often limited. Therefore, it is
unlikely that such a dedicated instrument will be built in the near
future.

In this article we advocate a different approach. By using an existing
spectrograph and only building an interface between it and an
interferometer at the same site, the combination of high spectral and
spatial resolution could be achieved on a much shorter timescale, and
for a fraction of the cost of a complete new instrument.

The pre-existing infrastructure would need to consist of two
telescopes, delay lines for path compensation, a fringe sensing unit
to acquire and stabilize the fringes, and a high-resolution
spectrograph on the same site. These conditions are already fulfilled,
or will be fulfilled in the very near future, at several
observatories. The two most promising sites are:
\begin{itemize}

  \item In the Southern hemisphere at Paranal Observatory with the
    VLTI in combination with the UVES spectrograph at Unit Telescope 2
    (UT2), or the High-Resolution IR Echelle Spectrometer (CRIRES)
    spectrograph at UT1.

  \item In the Northern hemisphere at Mauna Kea with the Keck
    Interferometer (KI) and HIgh Resolution Echelle Spectrometer
    (HIRES) at Keck I telescope or the NIRSPEC
    spectrograph at Keck II. Also at Mauna Kea, the OHANA (Optical
    Hawaiian Array for Nanoradian Astronomy) interferometer is
    currently under development, and will allow the combination of
    several other pairs of telescopes at the Manual Kea site.
\end{itemize}
In each case, three additional hardware components would be required:

\begin{enumerate}

\item A beam combiner that accepts two input beams from the telescopes
  and feeds the outputs carrying the fringe signals (coded as
  intensity variations) into fiber feeds;

\item Fibers that connect the interferometer to the spectrograph;

\item A fiber head that feeds the light from the fibers into the
  spectrograph.

\end{enumerate}

If separate telescopes are available for interferometry (as in the
case of the Auxiliary Telescopes of the VLTI), the impact on the
single-telescope-use of the spectrograph would be minimal, as it could
be used in the interferometric mode during times when other
instruments are scheduled for use with the main telescope.

The outline of this article is as follows. In
Section~\ref{sect:science} the scientific motivation for the proposed
setup will be discussed.  Section~\ref{instrument} will give an
overview of the challenges in designing and building the proposed
instrument, and their possible solutions. In
Section~\ref{sect:test_case:_vlti_uves} we will give more information
about the proposed combination of the VLTI with the UVES spectrograph,
and its expected performance. Partial and preliminary results of this
work have been presented in \cite{quirrenbach2008}. Section
\ref{sect:other_interferometer-spectrograph_pairings} highlights the
main points for other possible interferometer-spectrograph pairings,
and Section~\ref{Conclusion} gives our conclusions.

\section{Scientific case}
\label{sect:science}

By taking advantage of the existing infrastructure and instrumentation
at the observatories, the proposed interface between a high-resolution
spectrograph and an optical / near-infrared interferometer can provide
some unparalleled capabilities in a time- and cost-efficient
manner. Most importantly, interferometry can be performed with
sufficiently high spectral resolution to resolve absorption lines
allowing visibility changes across spectral lines to be measured even
in late-type stars. The interferometric spectra would also cover a
wide wavelength band. Using the broad spectral coverage, one would be
able to use cross-correlation techniques to obtain very accurate
radial-velocities and line shapes, which have proven extremely helpful
in planet search programs and asteroseismology.

Although restricted to observations of relatively bright stars,
interferometry at high spectral resolution will provide hitherto
inaccessible information on stellar rotation properties, atmospheric
structure and surface features, and can have a profound impact on a
large number of open questions in stellar astrophysics.

\subsection{Stellar diameters and limb-darkening}

Measuring the variation of the stellar diameter with wavelength, or
even better determining wavelength-dependent limb darkening profiles,
can provide a sensitive probe for the structure of strongly-extended
atmospheres of cool giant stars. Such data can be directly compared
with predictions of theoretical models, and provide qualitative new
tests of state-of-the-art three-dimensional stellar model atmospheres
\citep{quirrenbach2003}. These models make predictions for the
emergent spectrum at every point of the stellar disk. To compare model
predictions with data from traditional spectroscopy, they have to be
integrated over the full disk first. In contrast, interferometric
spectroscopy gives access to the center-to-limb variation of the
emergent spectrum, and is thus naturally suited to comparisons with
model atmospheres.

First steps in this direction has been made with the Mark III and COAST
interferometers and with aperture masking, by measuring the diameters of a
sample of cool giant stars in filters centered on deep TiO absorption bands and
filters in the nearby continuum \citep{quirrenbach1993, tuthill1999,
quirrenbach2001, young2003}. Many stars are found to be substantially larger in
the TiO bands, and to have wavelength-dependent asymmetry. It is easy to
understand the principle behind these effects: we effectively measure the
diameter of the $\tau$ = 1 surface of the star (taking limb darkening into
account properly), and the height of that surface varies with opacity and
therefore with wavelength. In cool stars this variation may be so large (up to
$\sim 10$\% of the stellar diameter for ``normal'' giants, even more for
pulsating variables) that it can be observed as a variation of the effective
stellar diameter with wavelength. The higher parts of the atmospheres are
cooler, making the brightness distribution across the stellar surface in
absorption bands more sensitive to asymmetries in the temperature distribution.
The large spectral widths of the filters used for these interferometric
observations average over many TiO absorption lines with different strengths.
Interferometric high-resolution spectroscopy will provide much more detailed
information on the diameter and limb darkening profiles as a function of TiO
absorption depth, and thus substantially better constraints on the theoretical
models.

Out of the giant stars which have been observed, the variations of the
apparent diameter and limb darkening profile with wavelength are most
pronounced in Mira stars. The instrument proposed here will enable
detailed investigations of the pulsation and wind acceleration
mechanisms. Again, high spectral resolution is required to sample a
large range of depths in the stellar atmosphere. The advantage of
combining high spatial and spectral resolution together in one
observation of an object rather than using separate observations in
this context has also been pointed out by e.g.\ \cite{Wittkowski2006,
  Tsuji2006}.

\subsection{Interferometric Doppler Imaging}

Classical Doppler Imaging (DI) has been developed into a very powerful
tool \citep[e.g.,][]{rice2002, kochukhov2004}. This technique allows
mapping of the chemical and magnetic properties of stellar
photospheres with surprisingly small details. Up to now, line profiles
have been used for DI which are based on average atmospheric
structures; this can obviously only be an approximation, in particular
in regions of extreme abundance peculiarity. Tools are now available
to compute such stellar atmospheres more accurately
\citep[e.g.,][]{shulyak2004}, and a reduced abundance contrast between
spots and their surrounding is expected. Interferometric
high-resolution spectroscopy data will allow a direct check of the
models, because abundance analyses can be performed for individual
surface regions of prominent chemically peculiar stars. The same
approach is also applicable for other stars with inhomogeneous surface
properties, like active cool giant stars, as interferometry allows the
study of individual surface regions. The fact that interferometry can
isolate the active regions will partly compensate for the lower total
signal-to-noise compared to single-telescope spectra, which always
average over the whole stellar surface. The spectral resolution
  needed depends on the projected rotational velocity of the star ($v
  \sin i$) and is likely to be greater then $20\,000$
  \citep{rice2002}. The required SNR will vary considerably from case to case,
  as it depends on the contrast of the structure to be imaged, on the number of
  lines that can be used for the analysis, and on the amount of any a priori
  knowledge available for DI.

\subsection{Pulsations and asteroseismology}

Radial and non-radial stellar oscillations lead to characteristic surface
patterns of line shapes and central velocities. The reconstruction of these
patterns from line profile variations alone is plagued with ambiguities,
however. These can to a large extent be resolved by the additional phase
information contained in interferometric data \citep{Jankov2001}. This means
that pulsation modes can be identified uniquely without any need for
comparisons with theoretical models. Empirical mode identification with
interferometric spectroscopy could become an important tool in the field of
asteroseismology \citep{cunha2007}.  A spectral resolution of the
  order of $40\,000$ will be needed with SNR per spectral
  resolution element of order 100 to 200 \citep{Jankov2001}. Observations with
  lower SNR may still be used in conjunction with cross correlation techniques
  \cite[e.g.][]{Queloz1995}. Spatially resolved observations may also provide
  access to higher-order ($l \geq 3$) modes, which have very small amplitudes
  and therefore normally not detectable in disk-averaged data.

\subsection{Interpretation of radial-velocity variations}

Radial-velocity observations of main-sequence stars have yielded more
than 300 planet detections so far. The wealth of information from
these surveys has revolutionized the field of planetary system
physics, but little is known about the incidence of planets around
stars with masses higher than about $1.5$~M$_\odot$, because more
massive main-sequence stars are difficult targets for radial-velocity
observations. Surveys of K giants can provide this information, and
planets around such stars have indeed been detected
\citep{frink2002}. Planets in highly eccentric orbits can be easily
identified as such in high-precision radial-velocity data due to the
distinct shape of the Keplerian velocity variations, but sinusoidal
variations observed in a number of objects in ongoing radial-velocity
surveys of K giants could be due either to planetary companions or to
low-order non-radial g-mode pulsations. It is possible in principle to
distinguish between these possibilities by analyzing the line shapes
(which should vary along with the radial velocity in the case of
pulsations, but remain stable in the case of companions), but this
requires very high spectral resolution and
signal-to-noise. Observations with interferometric spectroscopy could
resolve the stellar disk and hence distinguish more easily between
these possibilities, which would help to establish the mass function
of planets around stars with masses between $3$ and
$5$~M$_\odot$. Similar arguments apply to other cases in which
radial-velocity variations could plausibly be attributed to different
mechanisms, either related to stellar variability or to companions.

\subsection{Cepheids and distance ladder}

Limb darkening curves measured for a spectral line can provide direct
measurements of the projection factors of Cepheid pulsations, which
relate the true velocity of the pulsation to the observed
radial-velocity curve.  Uncertainties in this ``p factor'', which
presently must be computed from theoretical models, are a serious
limiting factor in current estimates of Cepheid distances with the
Baade-Wesselink method \citep{sabbey1995, marengo2002,
  Nardetto2006}. Interferometric spectroscopy can thus eliminate one
of the important contributions to the error budget for distances to
Cepheids and other variable stars.

\subsection{Orientation of stellar rotation axes}

Stellar rotation induces a difference in the fringe phase between the
red wings and the blue wings of stellar absorption lines in resolved
interferometric observations. Measuring the position angle of the
phase gradient allows determination of the orientation of the stellar
axis on the sky \citep{petrov1989, chelli1995}. More detailed modeling
of the interferometric signal can also provide the inclination of the
stellar rotation axis \citep{souza2004}. High resolution spectroscopy
will thus open a way to determine the orientation of stellar
rotational axes in space.

To know the orientation of the stellar axes in space is of particular
interest in double- or multiple-star systems. One can determine
whether the rotation axes of binaries are aligned with each other, and
with the orbital rotation axes of the systems. The orientation of the
rotational axes contains information about the origin and evolution of
the system. Such measurements have only be done for close eclipsing
binary systems as one can take advantage of the Rossiter--McLaughlin
effect during eclipses \citep[e.g.,][]{McLaughlin1924, rossiter1924,
  albrecht2007, albrecht2009}. However as short-period systems are synchronized
early in their life it is important to measure the orientation axes
for systems with larger semi-major axes, as they are of fundamental
importance for theories of binary star formation. Here the probability
that one can observe a eclipse and take advantage of the
Rossiter--McLaughlin effect is small.

One can also search for (partial) alignment of rotation axes in star
forming regions and stellar clusters. Interactions between stars in
multiple systems and other stars in a stellar cluster change the
angular momentum of the systems.

The orientation of the stellar rotation axes will be of special
interest for stars which harbor planets, since the mutual inclination
between the orbital plane of the companion and the rotation axis of
the star can provide insights into the formation and evolution
processes of the planet. If the orbital evolution of planetary systems
is dominated by few-body scattering processes, or the Kozai mechanism,
one might expect to find orbits that are not aligned with the stellar
angular momentum \citep[e.g.,][]{lin1997, Papaloizou2001, wu2003,
  nagasawa2008}.  The spin-orbit alignment has been measured in a
small number of systems with transiting extrasolar planets
\citep[e.g.,][]{winn2005}. Recently, the first planetary system with a
likely misalignment, XO-3, has been found \citep{hebrard2008}. As for
the binary systems these measurements are restricted to systems with
short orbital periods.  In the near future, astrometric orbits will
become available from ground-based and space-based astrometry with the
Phase-Referenced Imaging and Micro-arcsecond Astrometry (PRIMA)
facility at the VLTI, the Space Interferometry Mission (SIM), and with
GAIA. Combining this information with interferometric high resolution
observations will provide the relative inclination between the orbital
and equatorial plane for a large number of planets in a variety of
orbits.

\subsection{Differential stellar rotation}

Along with the oscillation spectrum, differential rotation is a
powerful diagnostic for the interior structure of a
star. Unfortunately, observations of differential rotation are
difficult with classical spectroscopy, and degeneracies exist between
inclination, limb darkening, and differential rotation
\cite[e.g.,][]{Gray1977}. These degeneracies can be resolved by the
additional information from interferometric spectroscopy
\citep{souza2004}.  High spectral resolution is of the essence for
this application, giving the proposed type of instrument a decisive
advantage over more conventional interferometric spectrographs such as
AMBER.

\subsection{Circumstellar matter}
\label{sect:circstellar_matter}

Velocity-resolved interferometric observations of emission lines can
be used to determine the structure and velocity field of disks around
pre-main-sequence objects and Be stars
\citep[e.g.][]{quirrenbach1997, young2003, tycner2006,
  meilland2007a}. It is possible to determine the disk opening angle
and the rotation law in the disk, to measure the location of the inner
edge of the disk, and to obtain detailed information on possible
asymmetries caused by spiral waves. Interferometric observations of
winds and outflows from pre-main-sequence stars and from evolved
objects can be used to determine their extent and overall geometry,
and to probe sub-structure such as clumps and shells. These
observations will not need the full spectral resolution offered by
high-resolution optical spectrographs, but access to the H$\alpha$
line is of critical importance.

\section{Instrument and infrastructure}
\label{instrument}

\begin{figure}
  \begin{center}
      \includegraphics[width=0.8\textwidth]{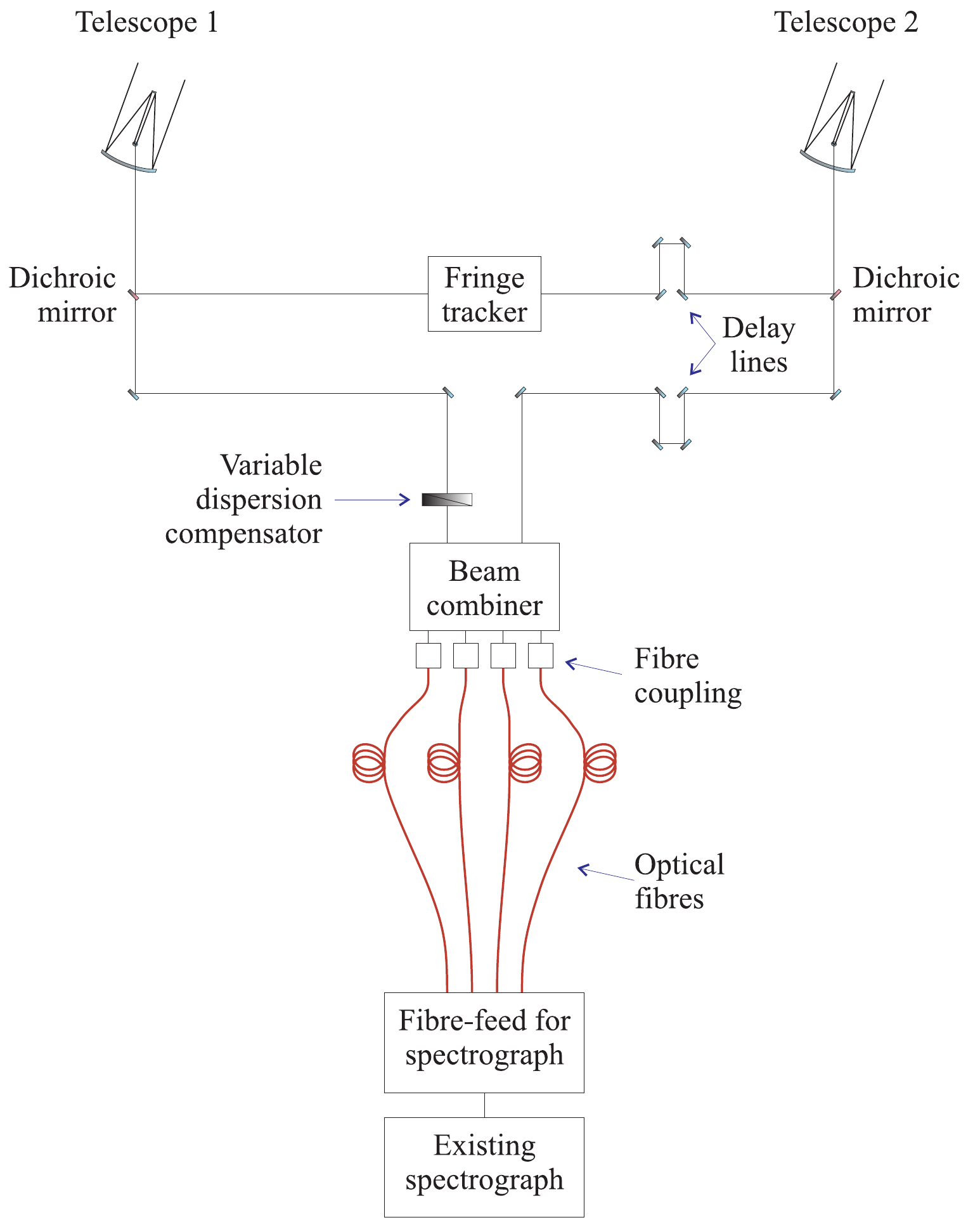}
      \caption{\label{fig:schematic_infrastrucutre} Schematic of the
        combined instrument, interferometer and spectrograph. The
        proposed instrument relies on pre-existing infrastructure
        (telescopes, fringe-tracker, delay lines and a spectrograph
        located within a few hundred meters). The additional
        components which must be built include the variable dispersion
        compensator, beam combiner, fiber coupling, optical fibers and
        fiber-feed for the spectrograph.}
  \end{center}
\end{figure}

\begin{figure}
  \begin{center}
    \includegraphics[width=0.8\textwidth]{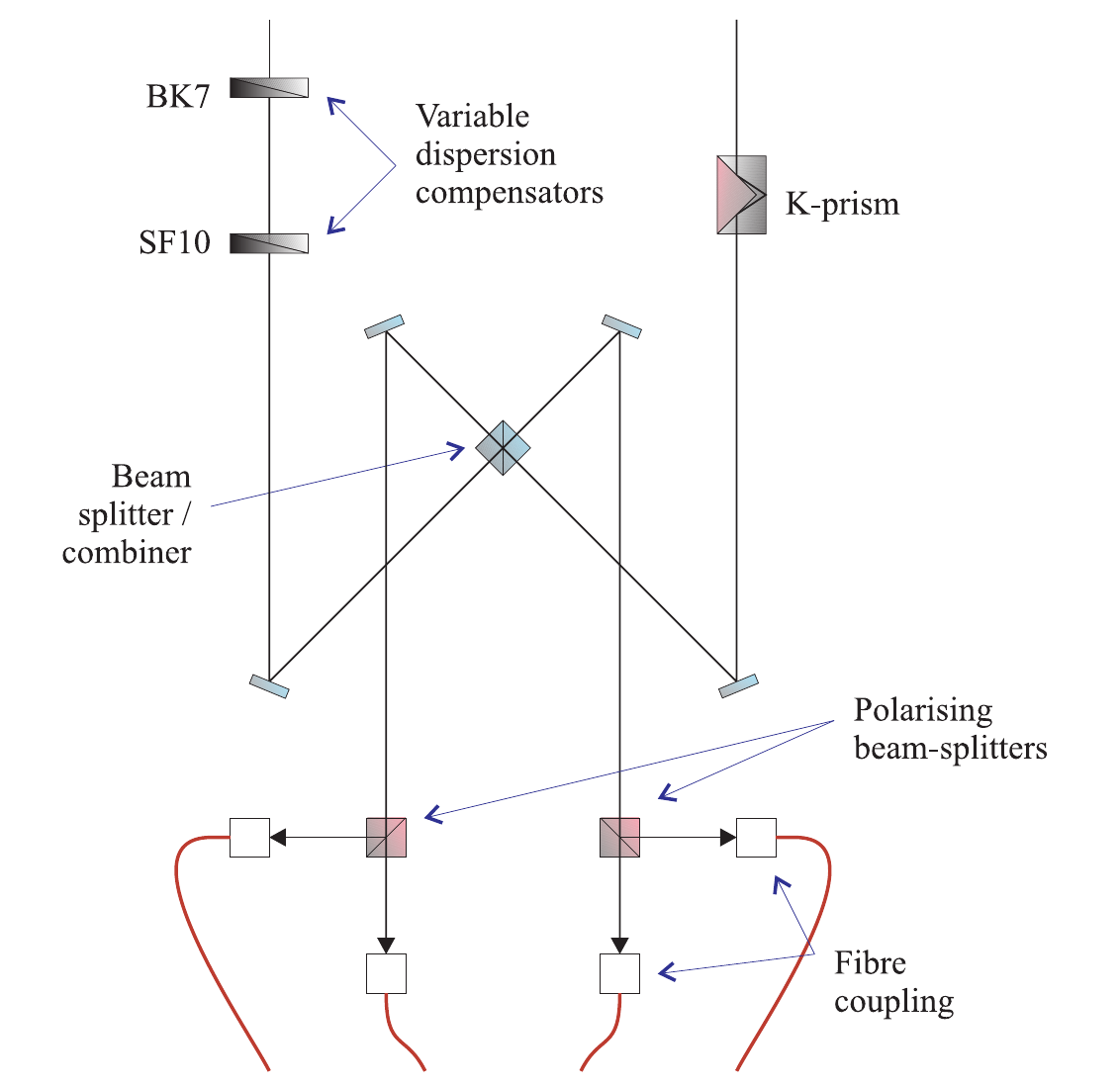}
    \caption{\label{fig:schematic_beam_combiner} Schematic of the
      suggested beam combiner for the coupling of a spectrograph to an
      interferometer. The beam from one telescope passes through a
      K-prism which introduces an achromatic phase shift of $\pi/2$
      between the s and p polarizations. The beam from the other
      telescope passes through longitudinal dispersion
      compensators. The combination of the two beams in the
      beam-splitter introduces a phase shift of $\pi$ between the two
      output beams. The two polarizations of the two output beams are
      separated by the polarizing beam-splitters, resulting in four
      beams with phase relations of $0$, $\frac{\pi}{2}$, $\pi$,
      $\frac{3\pi}{2}$ entering the fibers.}
  \end{center}
\end{figure}

In this section we describe our proposal for combining an optical/IR
interferometer with a high-resolution spectrograph.
Figure~\ref{fig:schematic_infrastrucutre} shows a schematic drawing of
the combined instrument, while a possible design for the beam combiner
is shown in Figure~\ref{fig:schematic_beam_combiner}. Special attention
is given to longitudinal dispersion compensation (Section
\ref{sect:dispersion_compensation}). This has to be addressed if one
wants to perform long integrations (several minutes) over a wide
bandpass, one of the key abilities of the proposed instrument.

\subsection{Telescopes}
\label{sect:telescopes_ao}

The Earth's atmosphere distorts planar incoming wavefronts from an
unresolved astronomical source, introducing phase fluctuations as a
function of position and time \citep{roddier1981}. These fluctuations
are commonly described by a Kolmogorov spectrum
\citep{tatarski61,kolmogorov41b} with constant Fried parameter $r_{0}$
\citep{fried66}. In order to obtain stable complex visibilities, the
incoming stellar wavefronts at the beam combiner have to be flat (with
constant wavefront phase as a function of position in the pupil plane
for an unresolved star). This can be achieved using either:
\begin{enumerate}
\item telescope aperture diameters small enough that the
  atmospherically-induced phase variations are negligible with
  $D/r_{0} \le 1$ (this may require a variable pupil stop);
 \item tip-tilt correction making wavefront errors negligible on
  aperture diameters up to $D/r_{0} \simeq 3$; or
\item higher-order adaptive optics.
\end{enumerate}
Larger apertures can be used at longer wavelengths or under better
seeing conditions, as the Fried parameter varies as $r_{0}\propto
\lambda^{6/5}$ and $r_{0}\propto 1/\mbox{FWHM}_{seeing}$ for
observations at a wavelength $\lambda$ and with a seeing disk of
FWHM$_{seeing}$.

\subsection{Longitudinal dispersion compensation}
\label{sect:dispersion_compensation}

To compensate for the difference in path length from the two
telescopes to the source, optical path must be added to one arm of the
interferometer. This is typically achieved through the use of
\emph{optical trombone} delay lines or through the stretching of
optical fibers \citep{monnier2003}. If the optical delay compensation
is performed in a dispersive medium, the Optical Path Difference (OPD)
where the fringes are found will vary with wavelength.  Longitudinal
dispersion can affect the OPD in the proposed instrument concept in
two ways:
\begin{enumerate}
\item Dispersion will give a variation of optical delay across the
  wavelength range of the spectrograph;
\item If the spectrograph operates in a different waveband from the fringe
  tracker, dispersion will introduce a different OPD in the spectrograph
  waveband to that in the fringe-tracking waveband.
\end{enumerate}

For a typical high-resolution spectrograph the coherence length of the
fringes in each spectral channel is much larger than the optical delay
offsets introduced by dispersion. However, time variation of the
dispersion during a spectrograph detector integration will blur the
interference fringes, reducing the measured visibility amplitude. The
two principal sources of variation in optical delay due to dispersion
are: \newcounter{list_change_on_spectrograph}
\newcounter{list_change_wrt_fringe_tracker}
\renewcommand{\labelenumi}{\Roman{enumi}}

\begin{SCfigure}
  \includegraphics[width=0.7\textwidth]{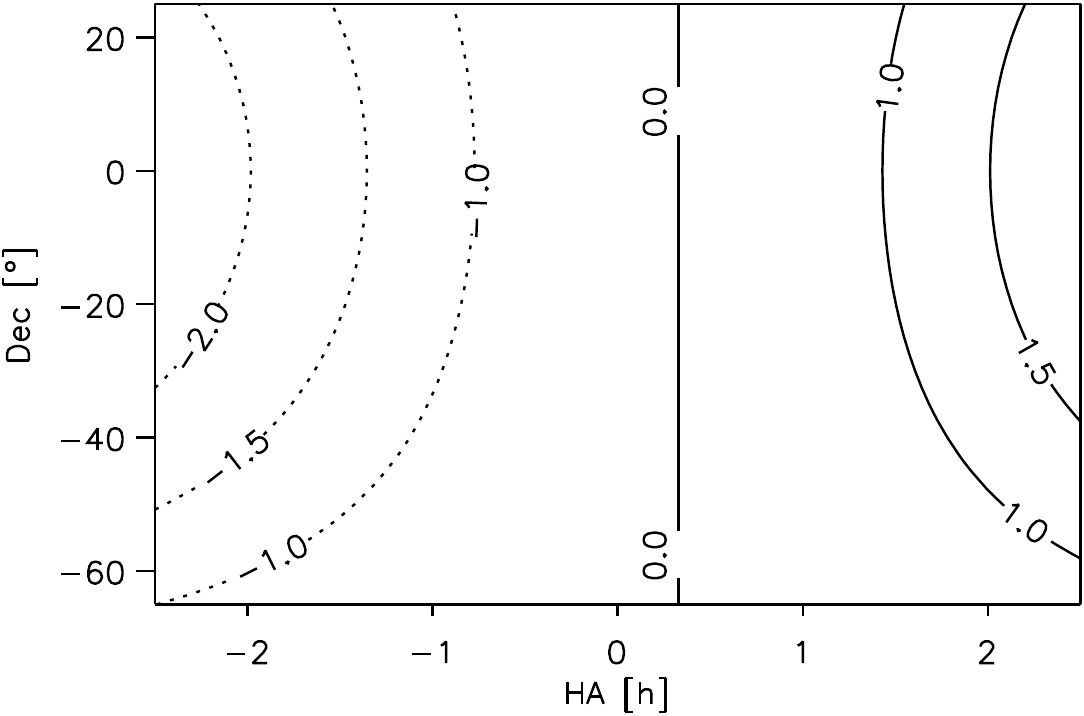}
  \caption{\label{fig:opd1} Change (in m) of OPD during a 10 minute
    integration as a function of HA and Declination for the G1-J6 AT
    stations at the VLTI which form a baseline of a length of 192\,m
    and an orientation angle of $-2^{\circ}$. An angle of $0^{\circ}$
    would indicate a baseline directed towards the North.}
\end{SCfigure}

\begin{SCfigure}
  \includegraphics[width=0.7\textwidth]{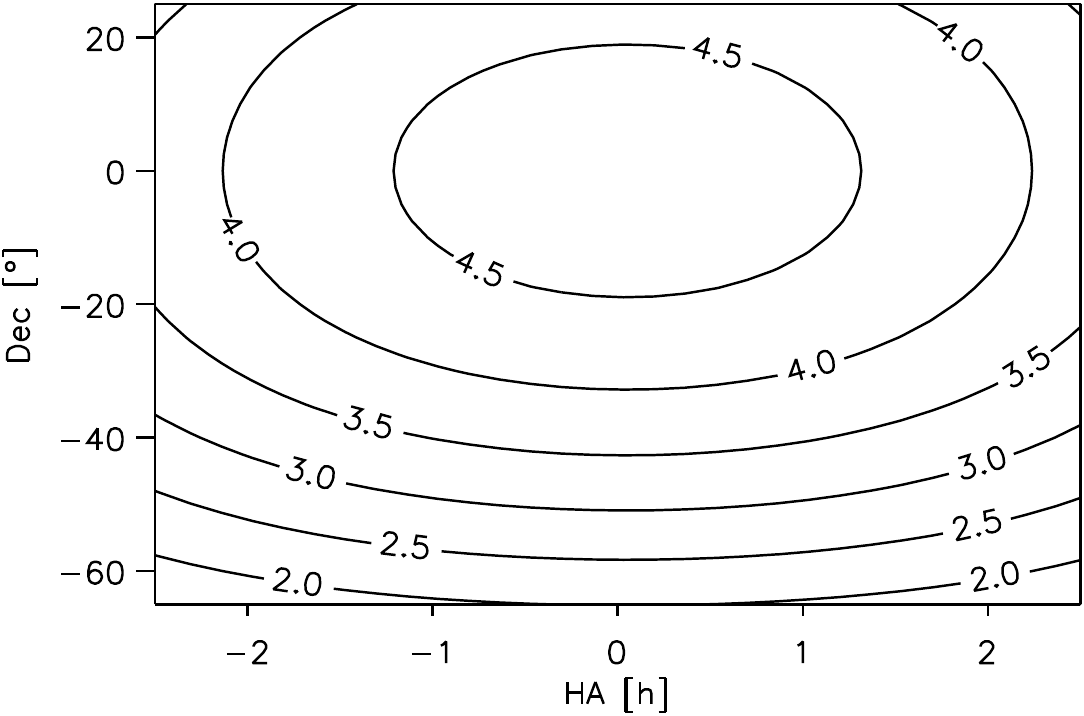}
  \caption{\label{fig:opd2} Same as Figure~\ref{fig:opd1}, but this time
    with a baseline of 109\,m length and an angle of 88$^{\circ}$
    (C1-J2).}
\end{SCfigure}

\begin{enumerate}
\item During an observation the position of the object on the sky
  changes. To keep the interference fringes at a stable position, the
  additional optical path introduced in one arm of the interferometer
  must be varied as the Earth rotates. The change in optical path
  through the dispersive medium in one arm of the interferometer
  causes the visibility phase to vary differently at each wavelength,
  unless the delay variation is introduced with an evacuated delay
  line. During the course of an observation of several minutes this
  inevitably leads to a loss of fringe contrast for observations in a
  waveband of non-zero bandwidth. For example, during a $10$-minute
  visible-light integration, the geometric delay path can change by
  several meters (see Figure~\ref{fig:opd1} and
  Figure~\ref{fig:opd2}), which would lead to a relative OPD shift of
  several $\mu$m between R band and I band.
  \setcounter{list_change_on_spectrograph}{\arabic{enumi}}
\item A change in the temperature or humidity of the air in one of the
  optical paths to the star will introduce a change in the column
  density of air and/or water vapor. To first order, these changes
  will be corrected by the fringe tracking. The residuals are not
  expected to be large enough to give different delays for different
  spectral channels in the spectrograph bandpass. However, if the
  spectrograph is operating in a different waveband from the fringe
  tracker, the optical delay in the spectrograph waveband may differ
  from that in the fringe tracking waveband.
  \setcounter{list_change_wrt_fringe_tracker}{\arabic{enumi}}
\end{enumerate}
\renewcommand{\labelenumi}{\arabic{enumi}}

Both dispersion problems could be circumvented by restricting the
exposure time, but this would reduce the observational efficiency, and
in the read-noise limited regime it would reduce the limiting
magnitude. Reducing the spectral bandwidth would solve point
\Roman{list_change_on_spectrograph} above, but would limit the
spectral coverage of the observations. As one can see in
Figure~\ref{fig:opd1} and Figure~\ref{fig:opd2}, the change in OPD depends
on the declination of the star and on the alignment of the baseline
with the rotation axis of the earth. Therefore, restricting the
baseline geometry and restricting the selection of sources would also
circumvent point \Roman{list_change_on_spectrograph} above. This would
seriously reduce the usefulness of the proposed instrument.

A better way to address point \Roman{list_change_on_spectrograph}
above would be to equip the beam combiner with a variable atmospheric
dispersion compensator (see
e.g.\ Section~\ref{sect:dispersion_compensation_at_uves-i}).  This
dispersion compensator would correct for the differential dispersion
introduced by the few meters of additional path between the two
telescopes added during the course of the observations; it does not
need to correct for the full differential air path.

Point \Roman{list_change_wrt_fringe_tracker} can be addressed by
estimating the dispersion at the wavelength of the spectrograph using
measurements of the ambient environmental conditions
\citep{albrecht2004} and the variation of optical delay with
wavelength across the fringe-tracking bandpass. If the fringe tracker
cannot operate sufficiently far from the zero optical group-delay
point, then an additional delay line will be required in order to
provide a different geometrical delay for the spectrograph than is
used for light that is sent to the fringe tracker, as shown in
Figure~\ref{fig:schematic_infrastrucutre}.

\subsection{Fringe tracker}
\label{sect:fringe_tracker}

The proposed instrument scheme relies on the interferometer having a
fringe-tracking capability. The fringe tracking must keep the fringes
of the spectroscopic observation stable even if they are observed at a
different wavelength to the one used for fringe tracking. The fringes
in the spectroscopic instrument must be kept stable to a fraction of a
wavelength (typically $\sim 1$\,rad of visibility phase). A number of
existing interferometers already have fringe-tracking instruments
\citep{delplancke2004,colavita2004,mcalister2004}. The RMS noise (the
jitter) in the optical delay from the fringe tracking will cause a
reduction in the visibility amplitude by a factor $\gamma$:

\begin{equation}
\gamma=\exp \left ( - 2 \left ( \frac{\pi \sigma_{ d }}{\lambda} \right )^{2}
\right ) \label{eqn:visibility_drop_from_jitter}
\end{equation}

where $\sigma_{d}$ is the RMS variation in the optical delay
compensation in a spectral channel with wavelength $\lambda$ during a
detector integration.

 If the fringe tracking wavelength differs from the science
  wavelength, differential refraction in the Earth's atmosphere will cause
  an additional reduction in visibility amplitude for high zenith
  angles (airmass $\geq 1.4$) of the target star
  \citep{colavita1987, quirrenbach1994}. This effect depends on the dispersion
  between the wavelengths of fringe tracking and observation, and therefore
  increases strongly towards the blue. As the proposed instrument
  will be used only for observations at lower airmass we do not
  include this into our performance analysis in
  Sect.~\ref{sect:performance}. 

\subsection{Beam Combiner}
\label{sect:beam_combiner}

The primary observable in an interferometer is the complex visibility
(having amplitude and phase), proportional to the complex coherence
function of the radiation received by the two telescopes
\citep{quirrenbach2001}. The complex visibility can be derived in a
number of ways, for example using a fixed delay offset and measuring
the fringe signal as a function of wavelength in the spectrum
\citep{labeyrie1975}. Alternatively, to obtain the full information on
the complex visibility in each spectral channel, one can measure the
four fringe quadratures at each wavelength, i.e. measure the light
intensity with fringe phase offsets of $0$, $\pi/2$, $\pi$ and
$3\pi/2$ radians. The normalized light intensities as a function of
wavelength $\lambda$ in these four outputs are commonly called $A\left
(\lambda \right )$, $B\left (\lambda \right )$, $C\left (\lambda
\right )$, and $D\left (\lambda \right )$ respectively. The $A\left
(\lambda \right )$, $B\left (\lambda \right )$, $C\left (\lambda
\right )$, and $D\left (\lambda \right )$ outputs can be produced
using $50$\% beam-splitter(s) (providing a $\pi$ phase shift between
the output beams) and achromatic $\pi/2$ phase shifts in two of the
four output beams (see
e.g. Figure \ref{fig:schematic_beam_combiner}). The complex visibility
$V\left (\lambda \right )$ in the spectral channel at wavelength
$\lambda$ is then fully described by the four intensities $A\left
(\lambda \right )$, $B\left (\lambda \right )$, $C\left (\lambda
\right )$, and $D\left (\lambda \right )$:

\begin{eqnarray}
  V\left (\lambda \right )= 2 \cdot \frac{A\left (\lambda \right ) - C\left
    (\lambda \right )}{A\left (\lambda \right ) + B\left (\lambda
    \right )+C\left (\lambda \right )+D\left (\lambda \right
    )} \nonumber \\
{}+2i \cdot \frac{B\left (\lambda \right ) - D\left (\lambda \right
    )}{A\left (\lambda \right ) + B\left (\lambda \right ) + C\left
    (\lambda \right ) + D\left (\lambda \right )} ,
    \label{eqn:comp_vis_from_abcd}
\end{eqnarray}

where $i=\sqrt{-1}$.

The squared amplitude of this visibility estimate $|V\left (\lambda
\right )|^2$ and the argument (fringe phase) $\phi\left (\lambda
\right )$ are given by:

\begin{eqnarray}
\label{Vsqest} |V\left (\lambda \right )|^2=4 \cdot \frac{\left( A\left
(\lambda \right ) -
      C\left (\lambda \right ) \right)^2 + \left( B\left (\lambda
      \right ) - D\left (\lambda \right ) \right)^2 }{\left( A\left
      (\lambda \right ) + B\left (\lambda \right ) + C\left (\lambda
      \right ) + D\left (\lambda \right ) \right)^2} \;, \nonumber
      \\ \phi\left
      (\lambda \right ) = \arctan \frac {A\left (\lambda \right ) -
      C\left (\lambda \right )}{B\left (\lambda \right ) - D\left
      (\lambda \right )} \;.
\end{eqnarray}

Applying the fringe estimators on a wavelength-by-wavelength basis,
one can thus derive the complex visibility as a function of
$\lambda$.\footnote{Note that the estimator for $|V(\lambda)|^2$ in
  Eqn.~\ref{Vsqest} is biased --- $|V(\lambda)|^2$ will be
  over-estimated in the presence of noise. Slightly modified
  estimators can be used to give unbiased estimates of
  $|V(\lambda)|^2$ \citep[e.g.,][]{shao1988}.}

The absolute phase will usually be corrupted by turbulence in the
Earth's atmosphere, but differential phases can be measured between
adjacent spectral channels or between the fringe tracking wavelength
and one of the observed spectral channels. These differential phases
can provide very valuable observables, such as phase differences
between the red and blue wings of spectral lines. In sources which are
resolved in some spectral lines but which are un-resolved at continuum
wavelengths (e.g. Be stars, as discussed in
Section~\ref{sect:circstellar_matter}), the complex visibilities can
be used to make interferometric images of the structure in the
spectral lines, using the position of the unresolved continuum source
as the phase reference. In this context it is worth noting that
phenomena on scales much smaller than the `resolution limit'
$\lambda/B$ of the interferometer with baseline $B$ are accessible
with this technique, because differential phases with a precision of a
few degrees provide astrometric accuracy significantly higher than the
conventional (imaging) resolution limit.

\subsection{Connection to the spectrograph}
\label{sect:spectrograph}

The outputs from the beam combiner can be easily transported to the
spectrograph using multi-mode fiber optics. Only the light intensity
as a function of wavelength is of interest at this point (after beam
combination), so the additional optical path length and longitudinal
dispersion from the fiber are unimportant.

\begin{figure}
  \centering
  \includegraphics[width=0.6\textwidth]{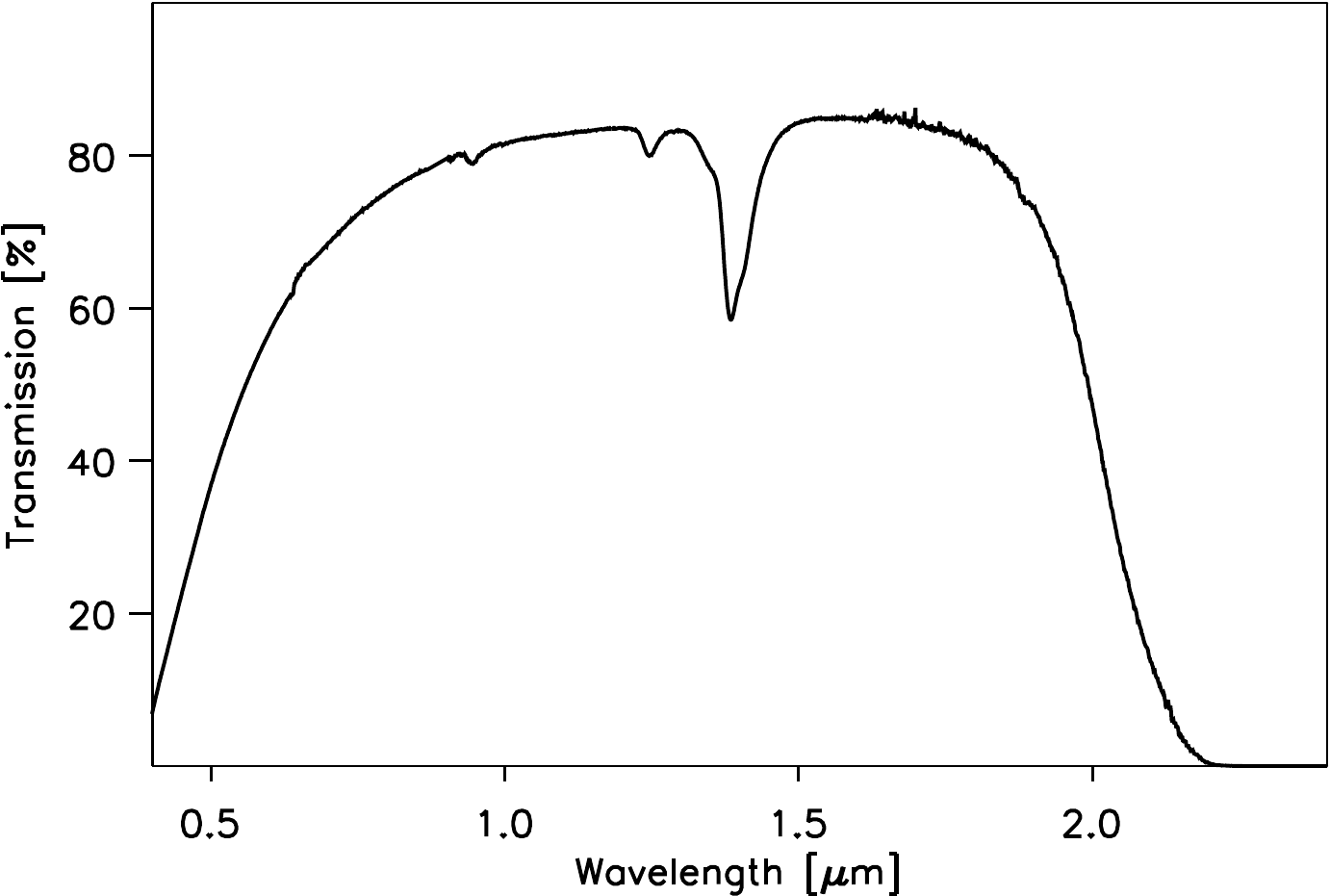}
  \caption{\label{fig:fiber_vis_nir} Transmission in the wavelength
 range from $0.4$ to $2.5$~$\mu$m for $150$\,m of Optran Plus WF fiber
 (Ceramoptec catalog, see http://www.ceramoptec.com/catalog.htm).}
\end{figure}

For a link at visible or near-infrared wavelengths one could for
example use the Optran Plus WF fiber with a core diameter of
$100$~$\mu$m (see Figure~\ref{fig:fiber_vis_nir}). Fluoride glass fibers
could provide better throughput at longer near-infrared wavelengths.

The insertion of pick-off mirrors might provide a good solution for
directing the light from the fibers into the spectrograph. However, in
some infrared slit spectrographs, the slit is located in a cryogenic
part of the instrument. If it is undesirable to make modifications
inside the cryogenic Dewar, it may be sufficient to place the fibers
in an image plane outside the spectrograph Dewar.

If the spectrograph is already a fiber-fed instrument, the number of
new components will be small and the installation fast.

\section{An illustrative example: UVES-I}
\label{sect:test_case:_vlti_uves}

In this section we investigate in more depth the possible combination
of the VLTI interferometer with the UVES spectrograph. We name this
combination \emph{UVES-I} in the reminder of this article. In particular
we address here the matter of external fringe tracking and dispersion
compensation in Sections \ref{sect:fringe_tracking_uves_i} and
\ref{sect:dispersion_compensation_at_uves-i}, respectively. As
mentioned above it is essential to solve these points if one wants to
combine a high-resolution Echelle spectrograph like UVES with an
optical interferometer like VLTI and carry out long exposures, which
has not been done so far.

\subsection{VLTI Auxiliary Telescopes}
\label{sect:telescopes_uves_i}

UVES-I would operate in the wavelength range between $0.6$~$\mu$m and
$1.0$~$\mu$m. Therefore it would use the VLTI ATs ($1.8$-m Auxiliary
Telescopes) and not the UTs ($8$-m Unit Telescopes) as the Multi
Application Curvature Adaptive Optics (MACAO) adaptive optics systems
of the UTs do not deliver well-corrected wavefronts in the visible.

The existing tip-tilt correction on the ATs would allow the use of
$3r_{0}$ sub-apertures with UVES-I (corresponding to $\sim 75$~cm at
$800$~nm under typical seeing conditions).  To minimize
fringe-tracking and tip-tilt correction errors, the PRIMA fringe
tracker, the tip-tilt system, and UVES-I should all use the same
aperture; a clear off-axis aperture could be chosen for this
purpose. The loss of two magnitudes in fringe tracking sensitivity
compared with the full AT apertures would be tolerable, as UVES-I
observations will usually target very bright stars. An arrangement in
which the fringe tracker and tip-tilt sensor use the full aperture,
and UVES-I the inner 75\,cm would provide higher fringe-tracking
sensitivity at the cost of increased phase noise in UVES-I. With the
planned installation of adaptive optics systems at the ATs, their full
1.8\,m apertures would become useable for UVES-I, corresponding to a
sensitivity gain of $\sim 2$\,mag.  Note that this estimate
  assumes a strehl ratio of $\sim0.5$ at $600$~nm, an ambitious
  value. 

The ATs currently have a dichroic beam-splitter sending the visible
light to the tip-tilt system, and passing the infrared light to the
delay lines and instruments. This beam-splitter reduces the VLTI
transmission in the visible considerably \citep{puech2006}. For UVES-I
this dichroic should be replaced by $10$-$90$ beam-splitter, with only
$10$\% of the visible light used for tip-tilt correction. The
astronomical targets of interest are all bright enough that they will
still give good tip-tilt performance. The sensitivity estimates given
in Section~\ref{sect:performance} are based on this change, and assume
a total VLTI transmission of $6$\% \citep{puech2006}.

\subsection{Fringe tracking with PRIMA}
\label{sect:fringe_tracking_uves_i}

Starlight at wavelengths longward of $1.5$~$\mu$m will be separated
using a dichroic mirror and sent to the PRIMA fringe tracker (see
Figure~\ref{fig:schematic_beam_combiner}) for stabilization of the
fringes. If the R-band fringes at the beam combiner can be stabilized
to less than one radian of fringe phase long integrations ($\gg$ than
the atmospheric coherence time) can be performed. In order to
stabilize the fringe phase at R-band, the OPD at R-band must be
calculated from the measured environmental conditions in the VLTI and
the measured phases in the PRIMA spectral channels
($1.95$--$2.45$~$\mu$m).

For the case of von Karman turbulence \citep{goodman85} with finite
outer scale $L_{0}$, the spatial structure function for the optical
phase $D_{\Phi}$ asymptotically approaches a maximum value of
$D_{\Phi} \left ( \infty \right )$ \citep{lucke07}:

\begin{equation}
D_{\Phi} \left ( r \right ) \equiv \left < \left | \Phi \left ( r' \right ) -
    \Phi \left ( r' + r \right ) \right |^{2} \right > \rightarrow D_{\Phi} \left ( \infty \right ) \mbox{, as } r \rightarrow \infty
\end{equation}

For von Karman turbulence, \cite{lucke07} give a possible range of:

\begin{equation}
0.0971 \left (\frac{L_{0}}{r_{0}} \right )^{5/3} < D_{\Phi} \left ( \infty \right ) < 0.173 \left (\frac{L_{0}}{r_{0}} \right )^{5/3}
\end{equation}

The total contribution of seeing to OPD fluctuations is typically
estimated by assuming that the seeing is caused by a wind-blown Taylor
screen of frozen turbulence passing the interferometer array
telescopes at a velocity $v$ \citep{taylor38, buscher95}. Under this
assumption, the asymptotic value of the temporal structure function
will be equal to the asymptotic value of the spatial structure
function:

\begin{equation}
D_{\Phi} \left ( t \right ) \equiv \left < \left | \Phi \left ( t' \right ) -
    \Phi \left ( t' + t \right ) \right |^{2} \right > \rightarrow D_{\Phi} \left (  \infty \right ) \mbox{, as } t \rightarrow \infty
\end{equation}

For a typical value of $r_{0}=0.25$~m at $800$~nm wavelength and
$L_{0}=22$~m \citep{martin2000} this would lead to a mean-square
optical phase variation $D_{\Phi} \left ( \infty \right )$ of:

\begin{equation}
170\mbox{~radians}^2 < D_{\Phi} \left ( \infty \right ) < 300\mbox{~radians}^2
\end{equation}

For a two-telescope interferometer with a long baseline, the
mean-square variation in fringe phase will be up to twice as large.
As this corresponds to fringe motions of several wavelengths, active
fringe tracking will be required in order to perform long
integrations.

The OPD fluctuations caused by astronomical seeing can be split up
into two constituent parts:
\begin{enumerate}
  \item variations in the mean particle density due to temperature
    fluctuations (temperature seeing, with no change in the air
    composition); and
  \item replacement of dry air with an equal particle density of water
    vapor (water vapor seeing, with no change in particle density).
\end{enumerate}

Measurements using the MIDI $10$~$\mu$m instrument at the VLTI
indicate that the differential column density of water vapor typically
varies by an RMS of $\leq 1$~mole~m$^{-2}$ (Meisner 2007, unpublished)
due to water vapor seeing (the displacement of dry air by an equal
particle density of water vapor). This introduces a mean-square fringe
phase fluctuation of up to 50\,rad$^2$ at $800$~nm wavelength. The
remaining $290$--$600$\,rad$^2$ of mean square phase variation result
from temperature seeing\footnote{A small fraction of the water vapor
  column density fluctuations is expected to occur within the VLTI,
  but this can be ignored when estimating the atmospheric contribution
  to the temperature seeing.}.

\begin{figure}
  \centering
  \includegraphics[width=0.8\textwidth]{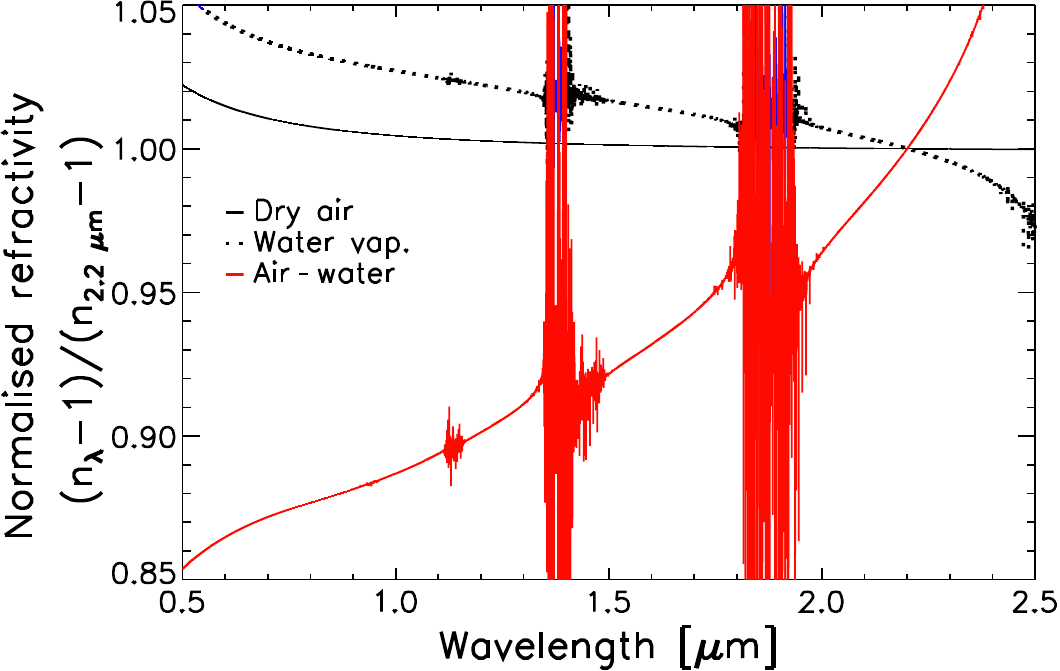}
  \caption{\label{fig:norm_refrac}Normalized refractivities of air and
    water vapor, plotted in black and blue respectively. The
    refractivity (equal to $n-1$ where $n$ is the refractive index) is
    plotted against wavelength for dry air and water vapor. The
    refractivity curves have been normalized in this plot such that
    the refractivity value at $2.2$~$\mu$m is unity. Also plotted in
    red is a curve of the refractivity of dry air minus the
    refractivity of water vapor, which has been normalized in the same
    way.  The normalization process stretches this curve
    vertically. The data were taken from \cite{Ciddor1996} and
    \cite{mathar2004}}
\end{figure}

If fringe tracking is performed at K-band, the fringes will be
partially stabilized in R-band and I-band. Figure~\ref{fig:norm_refrac}
shows plots of the refractivities of dry air and water vapor,
normalized to unity at a fringe-tracking wavelength of
$2.2$~$\mu$m. The black \emph{dry air} curve shows the relative amount
of OPD at each wavelength if temperature seeing in a dry-air
atmosphere introduced $1$~unit of OPD at $2.2$~$\mu$m. For air of
finite humidity, the plot of the relative amount of fringe motion
would be shifted linearly a small amount ($< 1\%$ of the way) towards
the blue \emph{water vapor} curve.

If the geometrical path length is varied in order to stabilize the
fringes at K-band against temperature seeing, and no additional
compensation is applied to account for the different refractivity in
the visible, the correction applied will be $0.5$--$1.5\%$ too small
to stabilize the fringes between $600$ and $900$~nm. For the
$288$--$602$~radians$^2$ of mean square phase variation expected from
temperature seeing, this will lead to an RMS fringe-tracking error of
$0.1$--$0.4$ radians at $800$~nm wavelength.

The red curve in Figure~\ref{fig:norm_refrac} shows the relative effect
of water vapor seeing at different wavelengths. It corresponds to the
difference in refractivity between air and water vapor of the same
(low) density, with the curve normalized to be unity at $2.2$~$\mu$m
(the fringe-tracking wavelength). A mean-square fringe phase
fluctuation of $\simeq 50$~radians$^2$ due to water vapor seeing at
$800$~nm wavelength will correspond to $\simeq 0.90$~$\mu$m RMS motion
at $800$~nm wavelength, but from Figure~\ref{fig:norm_refrac} it can be
seen that the fringe motion due to the water vapor seeing will always
be $15\%$ larger at $2.2$~$\mu$m. If K-band fringe tracking is
performed with no compensation for the different air refractivity in
the visible, this will introduce an RMS residual of $\simeq 1$~radian
to the fringe phase at $800$~nm wavelength. This would cause a
reduction of the fringe visibility due to blurring of the fringes in
long exposures.

The approach described by (Meisner 2007, unpublished) for N-band fringe
stabilization using PRI-MA can be adapted to improve R- and I-band
fringe stabilization during fringe phase tracking. In this method, the
geometric path from the star and the differential column density of
dry air are estimated from the position of the optical delay lines and
measurements from environmental sensors in the VLTI
\citep{albrecht2004}.  The K-band group phase measurements provided by
the PRIMA instrument are only very weakly dependent on the
differential column density of dry air, but are strongly dependent on
the differential column density of water vapor. If the effect of the
estimated dry-air column density is subtracted from the measured group
phase, the resulting group phase can provide a good estimate of the
water vapor column density. The fluctuations in the measured water
vapor column density can then be used to calculate and independently
correct the residuals produced by water vapor and dry air seeing,
assuming that the drift in the geometric delay error is small during
one integration.

\begin{SCfigure}
  \centering
  \includegraphics[width=7.2cm]{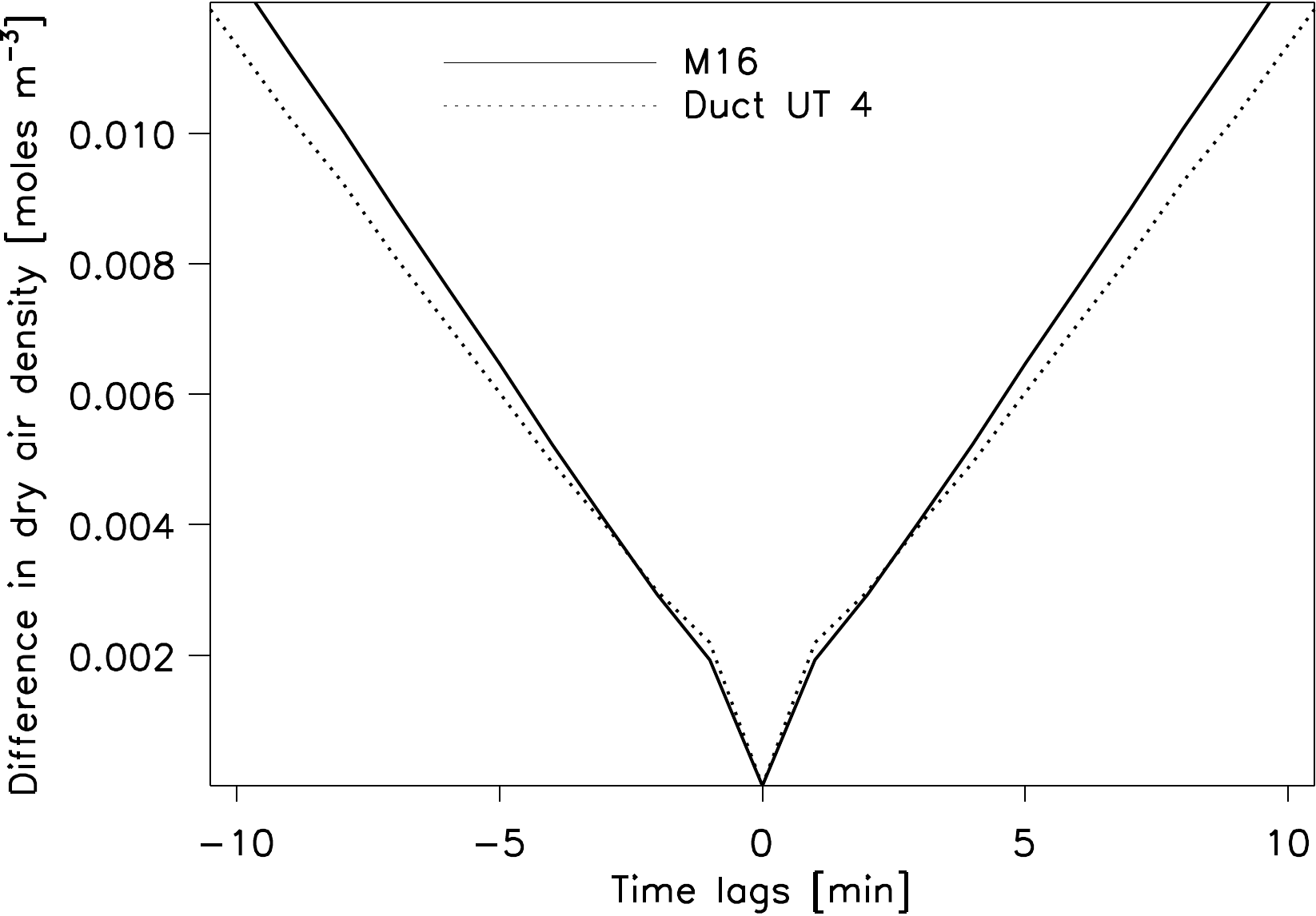}
  \caption{\label{fig:structure_func} Square root of the temporal
    structure function of the dry air density at two locations in the
    VLTI for the Night 22/23 September 2004. The solid line shows the
    temporal structure function for measurements at the M16 mirror in
    front of the interferometric laboratory, and the dashed line
    displays the same measurements inside the light duct towards the
    UT 4 telescope.}
\end{SCfigure}

Analyses of temperature, pressure and absolute humidity data taken on
4 nights immediately following the installation of the four humidity
and temperature sensors at Paranal \citep{albrecht2004} show that the
square root of the temporal structure function of the density of air
molecules typically varies by less than $0.003$ moles m$^{-3}$ (see
Figure~\ref{fig:structure_func}) at all measured locations within the
VLTI, during a 2-minute integration. There is no measurable
correlation between the fluctuations in the different ducts to the
telescopes and the fluctuations in the main delay tunnel on timescales
of a few minutes. The observed water vapor fluctuations within the
VLTI are less than $3\times10^{-4}$~moles~m$^{-3}$ over a 2-minute
integration, and are already included in the $\leq 1$~mole~m$^{-2}$
figure from (Meisner 2007, unpublished).

In the extremely pessimistic case that all the air in an entire
$100$-m duct simultaneously undergoes the same $0.003$~moles~m$^{-3}$
density change, $2$~$\mu$m of OPD would be introduced at a wavelength
of $2.2$~$\mu$m. The light paths at the VLTI pass through two such
ducts and through an air-filled main delay line, but the total OPD
fluctuation due to density changes within the VLTI is still expected
to be much smaller than the figure of $24$--$33$~$\mu$m RMS OPD
estimated above for temperature seeing in the atmosphere. The
measurements of K-band fringe phase and group phase include the
complete path through the atmosphere and instrument as far as the
PRIMA optical bench, so OPD fluctuations caused by changes in the
ambient conditions within the VLTI will be accurately tracked.

The expected fringe tracking noise introduced by the PRIMA instrument
itself will be $100$~nm for a star with $m_K = 8$, and $50$~nm at $m_K
= 6$ (see Figures~5-12 of \cite{delplancke2004}, and
\cite{tubbs2007}),  for a 75\,cm subaperture of the ATs these
  performance numbers are expected for stars that are two magnitudes
  brighter. Equation~\ref{eqn:visibility_drop_from_jitter} gives
visibility losses of $27$\% and $7.5$\% respectively at $800$~nm,
which is easily tolerable. PRIMA is currently being commissioned
  on Paranal, and the quoted fringe tracking noise for PRIMA will have
  to be verified during commissioning. Initial tests have already
  resulted in a fringe tracking rms of $\sim 120$\,nm, giving some
  confidence that a level of better than 100\,nm can indeed be
  achieved. It is also desirable that no fringe jumps occur during an
  exposure, as they introduce a delay offset corresponding to one
  wavelength in the fringe tracking band, and thus a reduction of the
  visibility in UVES-I. Occasional fringe jumps or fringe losses are
  not detrimental, however, if they are recognized and corrected
  quickly by the fringe tracker. For example, if the recovery time
  from such an event is one second, the visibility loss for a 200\,s
  exposure would be 1\% at worst, and it would vary slowly with
  wavelength.

\subsection{Dispersion compensation for UVES-I}
\label{sect:dispersion_compensation_at_uves-i}

Each meter of unbalanced air path introduces (under median Paranal
atmospheric conditions) offsets from the K-band group delay zero-point
of $108$~nm and $412$~nm, respectively, for the fringes at $900$~nm
and $600$~nm wavelength.  This fringe motion can be accurately
predicted from environmental sensor measurements and the known
geometrical delay. It can be stabilized at one wavelength using a
delay line, but to give good stabilization over the full $900$~nm to
$600$~nm wavelength range, a variable dispersion corrector is
required \citep[see][]{tango1990}.

Starlight at wavelengths shorter than $1.5$~$\mu$m is separated from
the K-band fringe tracking light by a dichroic mirror into the
dispersion compensator. The dispersion compensator and the beam
combiner are located on the same optical table as PRIMA to ensure
alignment and OPD stability (see
Figure~\ref{fig:schematic_beam_combiner}).

\begin{SCfigure}
  \centering
  \includegraphics[width=8cm]{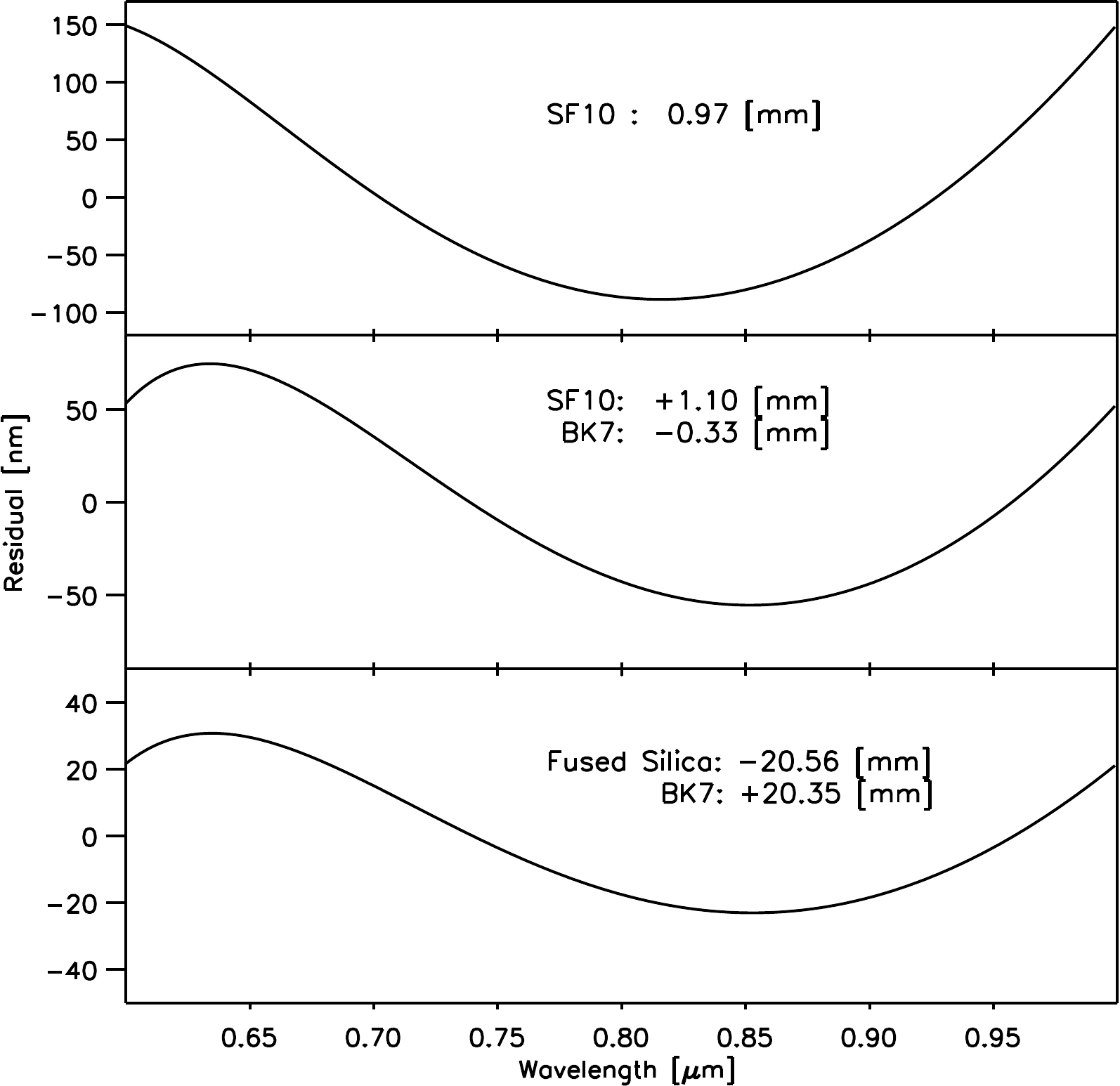}
  \caption{\label{fig:two_mat}Residual dispersion after accumulation
    of a $10$~$m$ change of the total delay, using atmospheric
    dispersion compensators made of: a] SF10; b] SF10/BK7; and c]
      BK7/Fused Silica. The combination of SF10 and BK7 (b]) could be
        used for an instrument like UVES-I.  The thickness of the
        material needed for the compensation is also indicated. For
        example an additional air path of $10$ m in one beam would
        require the insertion of a $0.97$\,mm SF10 glass in the other
        beam.}
\end{SCfigure}

Viable solutions for a single-material dispersion compensator
exist. Using data from the Schott catalog
(http://us.schott.com/sgt/english/products/listing.html), the Sellmeir
dispersion formula \cite[e.g.][]{Berger2003}, and Cramer's rule we
find that, for the $0.6$-$\mu$m to $1$-$\mu$m wavelength range SF10
glass can be used (see Figure~\ref{fig:two_mat}, upper panel). For
extreme parameters --- a build-up of $10$\,m delay during integration,
compare also to Figures~\ref{fig:opd1} and \ref{fig:opd2} --- the
residual uncompensated optical path length would reach $150$~nm,
however; therefore an arrangement where two different materials are
used for the dispersion compensation is preferred. A combination of
fused silica and BK7 glass gives the best theoretical performance, but
the required glass thickness is relatively high (see
Figure~\ref{fig:two_mat}, lower panel).  We therefore favor an SF10/BK7
combination, which provides compensation to $70$~nm even for extreme
delay rates, using much thinner glass elements (see
Figure~\ref{fig:two_mat}, middle panel). The variable dispersion
compensator will also compensate for the fixed additional dispersion
produced in the K-prism.

In addition to the compensation of the delay differences in the
visible band, we must also compensate the delay difference between the
visible band and the K-band, where PRIMA tracks the fringes. This can
be done by commanding the PRIMA fringe tracker to fringe track far
enough from the K-band group delay zero point to allow compensation of
the fringes at visible wavelengths, or alternatively by inserting
another additional delay as shown in
Figure~\ref{fig:schematic_infrastrucutre}.

\subsection{Beam combiner for UVES-I}
\label{sect:beam_combiner_uves_i}

One beam is sent through an achromatic phase shifter, which introduces a
$\pi/2$ phase shift of one polarization state with respect to the other. 
This phase shifter can be implemented with a K prism made of BK7, which
provides a nearly achromatic retardation of $\pi/2$ over most of the visible
wavelength range (see e.g.\ Bernhard Halle Product Catalog, www.b-halle.de).
Variable dispersion compensation is applied in the other beam. The dispersion
introduced by the K-prism is also compensated in the dispersion compensator.
Two flat mirrors then direct the beams into the main beam combiner
(Figure~\ref{fig:schematic_beam_combiner}). The combined beams, which have a
phase shift of $\pi$ relative to each other, are sent to polarizing
beam-splitters by a second pair of flat folding mirrors (see Figure
\ref{fig:schematic_beam_combiner}). The polarizing beam-splitters separate the
two linear polarizations (with their $\pi/2$ phase differences) giving the four
fringe quadratures $A\left (\lambda \right )$, $B\left (\lambda \right )$,
$C\left (\lambda \right )$, and $D\left (\lambda \right )$.

Full information on the complex visibility can be obtained from the
four fringe quadratures $A\left (\lambda \right )$, $B\left (\lambda
\right )$, $C\left (\lambda \right )$ and $D\left (\lambda \right )$
as discussed in Sect.~\ref{sect:beam_combiner}. For an ideal beam
combiner, the complex visibility can then be calculated for each
spectral channel using Eqn.~\ref{eqn:comp_vis_from_abcd}.

\begin{figure*}
  \centering
  \includegraphics[width=12cm]{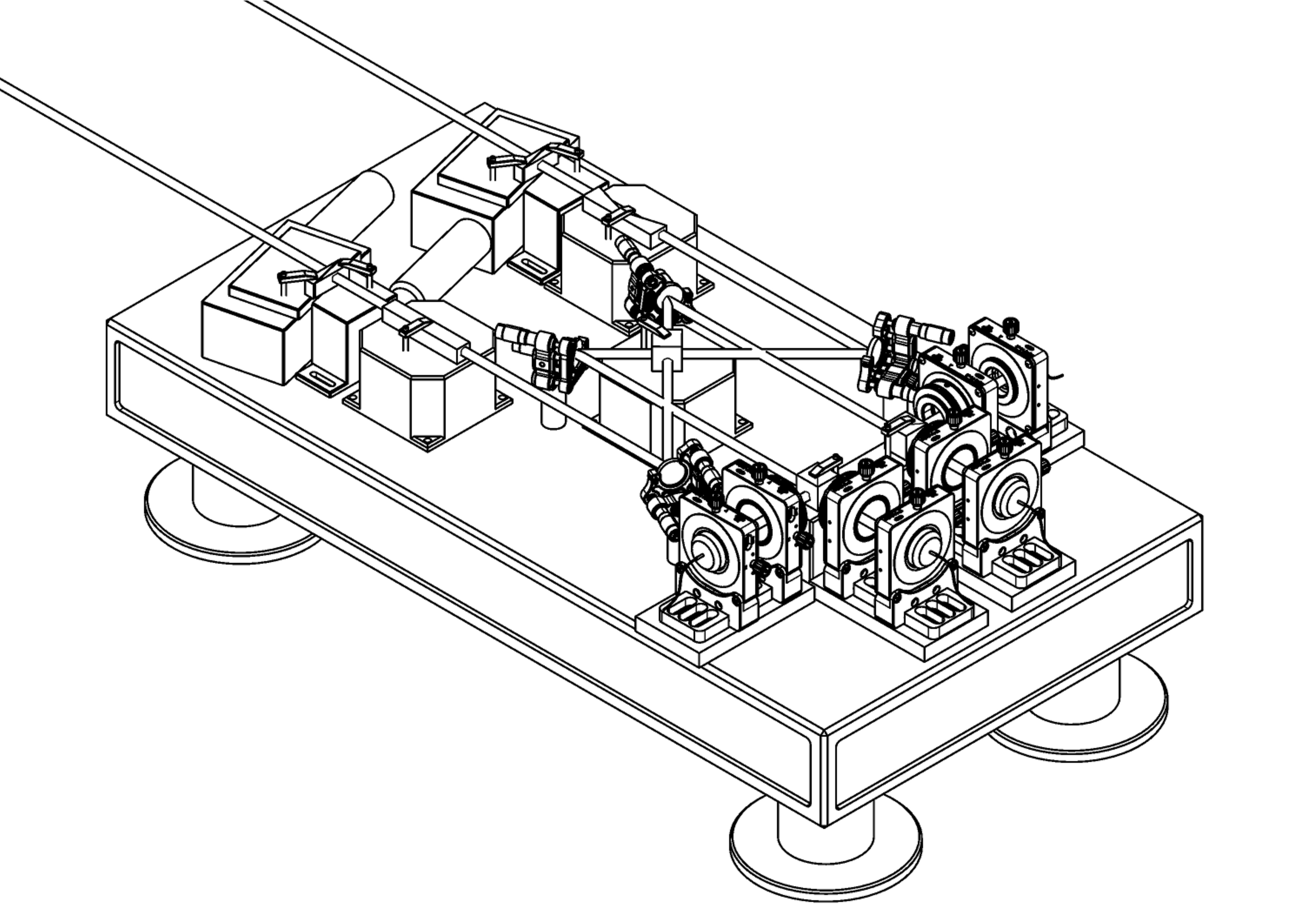}
  \caption{\label{fig:instrument} UVES-I beam combination table. The
    light beams from the two telescopes enter from the top left, and
    pass through the atmospheric dispersion compensator and achromatic
    phase shifter before being combined at the central beam
    combiner. The two polarization states are separated, and the
    resulting four beams coupled into optical fibers; the four fiber
    holders are visible at the bottom right. }
\end{figure*}

The requirements on the precision of the phase shift are rather loose;
one can easily tolerate up to $10^{\circ}$ error, even changing
  as function of wavelength, as long as they are stable. Errors in
the phase shifts can be compensated for in software, using a similar approach
to that used for phase measurements with the PRIMA instrument
\citep{tubbs2007}. Similarly, there are no stringent requirements on the
splitting ratios, their wavelength dependence, and the polarization
purity of the beam-splitter cubes. It is therefore possible to mostly use
standard commercial components; the use of a specialized high-performance
design for the non-polarizing beam splitter with $\lambda$-independent
near-50/50 splitting ratio and near-zero retardance between s- and
p-polarization as discussed e.g.\ by Shi et al.\ (2009) will hardly be
necessary. A preliminary design of the beam combiner fits on a
$45$\,cm\,$\times 85$\,cm breadboard (see Figure~\ref{fig:instrument}). This
compact design ensures stability and minimizes the space taken up in the
beam-combination laboratory of the interferometer. Due to its high spectral
resolution, the coherence length of UVES-I is very large ($\ge 3$\,cm).
However, the drift of the optical path difference \emph{during} one exposure
must not be larger than $\sim 50$\,nm, as this would reduce the fringe
visibility due to the fringe blurring; therefore the UVES-I breadboard should
be mounted as rigidly as possible to the fringe sensing unit. This eliminates
the necessity of an additional metrology system.

\subsection{UVES instrument on UT-2}
\label{sect:uves_spectrograph}

The impact of the proposed combination on the UVES instrument would be
small.  The UVES-I fiber interface to the spectrograph will be similar
to the existing link from FLAMES (Fiber Large Array Multi Element
Spectrograph) to UVES \citep{pasquini2000}, which has a fiber head for
$8$ fibers. The UVES-I fiber head will be placed after the pre-slit
optics and can, when UVES-I is used, be injected via a folding mirror.

To retrieve the full interferometric information only four fibers are
needed. It would thus be possible to use two VLTI baselines
simultaneously if an eight-fiber feed is used. This would require both
of the PRIMA fringe-tracking units to stabilize the fringes on the two
baselines. Interesting, but currently not possible at Paranal for an
instrument like UVES-I, is the simultaneous use of three baselines,
which would make it possible to measure a triangle of baselines, and
aquire closure phase. For this a three-way beam combiner, with 12
output fiber is needed and the light from these fibers has to be
directed on the spectrograph CCD with sufficiently-low cross-talk
between the different orders.

As the light is injected via fibers into the spectrograph the spectral
resolution achieved by the instrument no longer depends on the slit
size, but on the core diameter of the fiber and the re-imaging
optics. For example using a fiber with a core diameter of
$100$~$\mu$m, the fiber whose transmission is shown in
Figure~\ref{fig:fiber_vis_nir}, and with re-imaging optics changing the
F/3 fiber output beam into the F/10 beam similar to that produced by
the standard pre-slit optics of UVES \citep{Dekker2000}, a spectral
resolution of $\sim 55 000$ for UVES-I would be achieved, while still
making alignment of the fiber-couplers easy.

\subsection{Performance}
\label{sect:performance}

The detector integration times used should be sufficiently long to
ensure that the detector read-noise is not dominant (see below), and
sufficiently short to allow compensation of the atmospheric dispersion
accumulating during the exposure with a relatively simple device (see
Section \ref{sect:dispersion_compensation}). This will typically give
exposure times between $1$\,min and $15$\,min.

The beam combiner has $12$ optical surfaces (before fiber
coupling). The transmissive optics will be covered with
anti-reflection coatings optimized for the wavelength range used,
while mirrors will be coated in protected silver.  Using a
conservative assumption of $95$\% efficiency per mirror surface and
$99$\% efficiency per anti-reflection coated glass element, the
resulting overall throughput of the beam combination table is $72$\%.

The transmission of the fiber link depends on the fiber coupling
efficiency as well as the bulk transmission losses in the fiber. The
fiber link will be similar to the fiber link of FLAMES but the fiber
length will be larger, about $150$~m compared to $40$~m for the
FLAMES-UVES link. In Figure~\ref{fig:fiber_vis_nir} the transmission for
$150$~m of Optran Plus WF fiber is given. Note that the transmission
changes from $55$\% to $80$\% between $0.6$~$\mu$m and $1.0$~$\mu$m

Assuming two optical surfaces at each end of the fiber couplers plus
one folding mirror which can inject the light into UVES while the
spectrograph is in interferometric mode, and the transmission
efficiencies described above, the total transmission of the UVES-I
link will vary from $45$~\% to $65$~\% over the
$0.6$~$\mu$m--$1.0$~$\mu$m wavelength band used.

The system visibility of an interferometer, i.e., the fringe contrast
measured on an unresolved source, depends on the wavefront quality of
the two beams and on the accuracy of overlap at the beam
combiner\footnote{We do not consider the possibility of using larger apertures with a
spatial filter here, because this would reduce the temporal stability of
the fringes \citep{tubbs2005, Buscher2008} and the introduction of
such a filter would impose much more stringent requirements
on the alignment.}. Factors influencing the system
visibility include alignment, aberrations in the instrument optics,
tracking (tip-tilt) errors, and higher-order wavefront errors due to
atmospheric seeing. Using tip-tilt corrected $3r_{0}$ apertures we
expect a mean Strehl ratio of $\ge 43\%$ from each telescope
\citep{noll76, fusco2004}, providing a system visibility of order
$0.40$.

The total difference in sensitivity between FLAMES and the UVES-I
combination depends on the instrument efficiency and the source
visibility. The efficiency due to both the throughput and the system
visibility would be equivalent to a loss of $10.8$ magnitudes
(Tab.~\ref{table:uves_i_uves_comp}) in both the stellar and sky
background photon counts. After an integration of $2$~min using the
maximum spectral resolution and fast read-out rate of UVES-I, one
would be in the photon-noise-limited regime for a G0 star of magnitude
$R=6$. In this regime long integrations can be constructed from
multiple short exposures with no additional noise penalty in order to
reach the required SNR. The signal-to-noise also varies in proportion
to the source visibility.

Variations in seeing will influence the limiting magnitude of UVES-I
in two ways: The fringe-tracking performance will decrease with
increased seeing and the aperture will need to be stopped down.
Assuming somewhat worse than typical seeing conditions of $L_{0}=40$~m
and $r_{0}=0.15$~m at $656$~nm (compared to the values given in
Sect.~\ref{sect:fringe_tracking_uves_i}) the aperture would need to be
stopped down to $0.45$m to stay at $3$ r$_{0}$ ($1.1$ magnitudes).
Together with decreased fringe tracking performance ($0.2$ magnitudes
) the total loss of sensitivity of UVES-I would be of $12.1$
magnitudes relative to FLAMES.

It should be pointed out that even
though the sensitivity of UVES-I is significantly lower than that of
UVES, the final sensitivity compares quite favorably with other
optical interferometers especially bearing in mind that the achieved
spectral resolution is very high. The signal-to-noise ratio will be
further reduced in proportion to the source visibility, but for many
measurements described in Section~\ref{sect:science} the sources need
only be marginally resolved, so that the variation of visibility phase
with wavelength directly probes offsets in the position of the
emitting region.

\begin{table}
\begin{center}
\caption{Comparison of UVES-I throughput with the throughput of
  UVES-FLAMES, for operation without adaptive optics (0.75\,m
  aperture) and with adaptive optics (full 1.8\,m aperture of the
  ATs). \label{table:uves_i_uves_comp}}
\smallskip
\smallskip
\smallskip
\begin{tabular}{p{7cm}|c}
Term & Flux reduction ($\frac{\mathrm{UVES-I}}{\mathrm{FLAMES}}$) \\
     & \\
\hline
Telescope aperture area (no AO / with AO) & $0.009$ / $0.05$                   \\
Number of telescopes                & $2$                                      \\
Transmission VLTI / UT              & $0.075$                                  \\
Beam combiner transmission          & $0.72$                                   \\
Fiber link UVES-I / FLAMES          & $0.75$                                   \\
Flux for
real and imaginary components of
visibility                          & $0.5$                                    \\
Squared system visibility           & $0.16$                                   \\
Squared fringe tracking losses      & $0.85$                                   \\
 & \\
\textbf{Total (no AO / with AO)}    & $4.9\times10^{-5}$ / $2.8\times10^{-4}$  \\
                                    & $10.8 \mathrm{mag}$ / $8.9 \mathrm{mag}$ \\

\end{tabular}
\end{center}
\end{table}

For example, for a measurement of the phase change across a spectral
line the adjacent continuum provides an excellent calibration,
allowing measurement of phase changes with wavelength as small as one
milliradian. This means that UVES-I will be sensitive to astrometric
offsets that are over one hundred times smaller than the resolution
limit of the interferometer. It will not be possible to image such
small structures, but models of such astrometric offsets can be
constructed and tested with UVES-I.

It should be pointed out that not all observing programs will require
the full spectral resolution provided by the new instrument. In those
cases it will be possible to obtain a higher SNR per desired spectral
element by binning the data. In the readout-noise-limited regime, this
can best be done on the CCD.  This would have the advantage that one
would be able to go to fainter targets than stated above with shorter
integration times and shorter detector readout times.

Often the astrophysical interesting signal is encoded in multiple
lines in a spectral region (Section~\ref{sect:science}). In this case
the SNR can be improved by cross-correlating these absorption lines
with a line-template mask. This is routinely done in high-resolution
spectroscopy, \cite[e.g.][] {Griffin1967, Queloz1995, Donati1997,
  Rucinski1999} but has not yet been done with optical
interferometers, while it would be possible with an instrument like
UVES-I.

\section{Other interferometer-spectrograph pairings}
\label{sect:other_interferometer-spectrograph_pairings}

In this section we give some remarks about other possible pairings of
interferometers and spectrographs at different sites and in different
wavelength regimes.

The high-resolution spectrograph CRIRES at Paranal Observatory covers
the wavelength range from $1$~$\mu$m to $5$~$\mu$m and has a spectral
resolution up to $R_{CRIRES}\sim10^{5}$ \citep{kaeufl2004}. The
wavelength range would overlap with the wavelength range of AMBER
\citep{petrov2003} but with a much higher resolution
($R_{AMBER}\sim12 000$). The throughput of the VLTI+CRIRES
combination would be higher than UVES-I as the transmission of the
VLTI increases towards longer wavelengths. The distortions of the
incoming wavefronts are also less severe for the longer wavelengths
(see Section \ref{sect:telescopes_ao}), so one would be able to use
the full aperture of the ATs or make observations with the UTs using
the MACAO AO system \citep{arsenault2004}. For a combination of CRIRES
with the VLTI one could use the Optran Plus WF fiber
(Figure~\ref{fig:fiber_vis_nir}) as the transmission for a $150$~m of
Optran Plus WF fiber between $1$~$\mu$m~and~$2$~$\mu$m would be
$\sim$ $80$~\%. Note the dip at $\sim$~$2$~$\mu$m and the steep
decline in transmission beyond $2$~$\mu$m.

In the Northern hemisphere the Mauna Kea Observatory site would be
well-suited for the high spectral resolution optical/IR interferometer
design advocated in this article. The Keck interferometer (KI) is
already in place at the observatory and its infrastructure is similar
to the one found at the VLTI (delay lines, $2$-$\mu$m fringe
tracker). The HIRES spectrograph with the spectral range from
$0.3$~$\mu$m~to~$1.1$~$\mu$m and a resolving power of $R\sim67 000$
\citep{vogt1994} at the Keck I Telescope covers a similar
wavelength-range to the UVES spectrograph. Therefore a similar
approach as described in Section~\ref{sect:test_case:_vlti_uves} is
possible for mating HIRES and KI.  The combination of a
high-resolution spectrograph with an interferometer at this site seems
particularly attractive from the scientific point of view, as so far
the interferometer has no spectrographic mode.

A second interferometer, the OHANA experiment, is also currently being
set up at Mauna Kea \citep{perrin2004}. This instrument uses optical
fibers for beam transport instead of the bulk optics used in
conventional interferometers. This allows combination of telescopes
which were not originally designed for interferometry, forming an
array several hundred meters across. Using the OHANA array, one would
potentially have access to all spectrographs at the telescopes which
are connected to OHANA. However one has to use the bandwidth offered
by the OHANA fibers (J,H,K). One would prefer to keep the length of
the transport of the beams after the combination in the
interferometric lab short to avoid unnecessary losses, therefore one
could use the spectrographs at the telescope where the beam combiner
is placed. A new beam combiner at CFHT will be used by the OHANA
project in addition to the existing K-band fringe tracker on the Keck
Interferometer. The Near Infrared Echelle Spectrograph (NIRSPEC) at
the Keck II could be used as the spectrograph.

\section{Conclusion}
\label{Conclusion}

This article shows that the combination of high-resolution
spectroscopy with long-baseline interferometry gives access to
hitherto unobservable properties of stellar surfaces and circumstellar
matter.

Furthermore the article shows how this combination can be achieved
without building a major new instrument; only a beam combiner and a
fiber link are needed. No instrument components are required which
would be expensive or time-consuming to make. The use of external
fringe tracking is essential for an instrument like this as it enables
long integration times. Time-varying longitudinal dispersion could
severely limit the possibilities of such an instrument. Dispersion
compensation techniques are investigated in this article and a
solution is presented which allows integration times up to a few
minutes.

The implementation of this approach is shown for the example
combination UVES-VLTI. The resulting instrument would differ from
other instruments or efforts taken to achieve high spatial and
spectral resolution. It would offer spectral resolution nearly a
factor 2 higher than for any other interferometric instrument, and
over a wide spectral range of a few hundred rather than a few tens of
nanometers. It offers the same possibilities to an astronomer as a
high-resolution Echelle spectrograph behind a single telescope does,
plus the high spatial resolution due to the interferometer, albeit
only for bright targets. However it is important to realize that this
concept, the combination of two existing instruments to create a new
one, is not limited to a specific location or instrument, but rather
is an approach which can be followed at different observatories in
both hemispheres.

It is worth pointing out that the measurements taken with the proposed
instruments are differential in nature (e.g. change of visibility
amplitude and visibility phase over spectral lines), allowing many
interferometric calibration problems to be circumvented. 
  Therefore science relying on absolute V$^{2}$ measurements would
  need additional data form other instruments. However as noted in
Section~\ref{sect:science}, differential measurements contain a wealth
of astronomical information in the optical/IR regime.

\begin{acknowledgements}
  We are thankful to Jeff Meisner, Luca Pasquini, and Gerardo Avila
  for many useful discussions and suggestions. We are grateful to
  Richard J. Mathar for providing material on the refractivity of
  water vapor. We would like to thank the anonymous referee for
   her/his detailed and insightful comments, which improved the quality
    of the paper.
\end{acknowledgements}

\end{document}